\documentclass[useAMS,usenatbib]{mn2e}
\usepackage{graphicx}
 
\def\be{\begin{equation}} 
\def\ee{\end{equation}}

\def\HI{\hbox{H~$\scriptstyle\rm I\ $}}

\def\gsim{\lower.5ex\hbox{\gtsima}} 
\def\lsim{\lower.5ex\hbox{\ltsima}} \def\gtsima{$\; \buildrel > \over 
\sim \;$} \def\ltsima{$\; \buildrel < \over \sim \;$} \def\prosima{$\; 
\buildrel \propto \over \sim \;$} \def\gsim{\lower.5ex\hbox{\gtsima}} 
\def\lsim{\lower.5ex\hbox{\ltsima}} 
\def\simgt{\lower.5ex\hbox{\gtsima}} 
\def\simlt{\lower.5ex\hbox{\ltsima}} 
\def\simpr{\lower.5ex\hbox{\prosima}} \def\la{\lsim} \def\ga{\gsim}

\def\gtsima{$\; \buildrel > \over \sim \;$} 
\def\ltsima{$\; \buildrel < \over \sim \;$} 
\def\gsim{\lower.5ex\hbox{\gtsima}} 
\def\lsim{\lower.5ex\hbox{\ltsima}} 
\def\simgt{\lower.5ex\hbox{\gtsima}} 
\def\simlt{\lower.5ex\hbox{\ltsima}} 
\def\simpr{\lower.5ex\hbox{\prosima}} 
\def\la{\lsim} 
\def\ga{\gsim}

\def\E3{{\cal E}_{\rm g}^{III}}

\title[LAE sub-mm detection]{Detecting Lyman Alpha Emitters in the Sub-millimeter} 
\author[Dayal, Hirashita \& Ferrara]{Pratika Dayal$^{1}$\thanks{E-mail:dayal@sissa.it (PD)},
 Hiroyuki Hirashita$^{2}$ \& Andrea Ferrara$^{3}$  \\ 
$^{{1}}$ SISSA/International School for Advanced Studies, Via Beirut 2-4 Trieste, Italy, 34014\\ 
$^{2}$ Institute of Astronomy and Astrophysics, Academia Sinica, P.O. Box 23-141, Taipei 10617, Taiwan\\ 
$^{3}$ Scuola Normale Superiore, Piazza dei Cavalieri 7, 56126 Pisa, Italy}

\begin{document} 
 
%\date{Received 2008 December 24; in original form 2008 December 24} 
 
\pagerange{\pageref{firstpage}--\pageref{lastpage}} \pubyear{2009} 
 
\maketitle 
 
\label{firstpage} 
\begin{abstract} 
Using the results from a previously developed Ly$\alpha$/continuum production/transmission and dust enrichment model for Lyman Alpha Emitters (LAEs), based on cosmological SPH simulations, we assess the detectability of their dust-reprocessed sub-millimeter (submm) radiation. As supernovae (rather than evolved stars) control dust formation and destruction processes, LAEs are relatively dust-poor with respect to local galaxies: they have low dust-to-gas ratios (0.05 times the dust-to-gas ratio of the Milky Way) in spite of their relatively high metallicity, $Z \approx 0.1-0.5 Z_\odot$. Using the derived escape fraction of ultraviolet (UV) continuum photons we compute the UV luminosity absorbed by dust and re-emitted in the far infrared. The LAE submm fluxes correlate with their Ly$\alpha$ luminosity: about $(3\%,1\%)$ at $z=(5.7,6.6)$ of the LAEs in our simulated sample (those with with ${\rm Log}\, L_\alpha > 43.1$) would have fluxes at 850 $\mu$m (the optimal band for detection) in excess of $0.12$~mJy and will be therefore detectable at $5\sigma$ with ALMA with an integration time of only 1 hour. Such detections would open a new window on the physical conditions prevailing in these most distant galaxies.
\end{abstract}

\begin{keywords}
 methods:numerical - galaxies:high redshift - radiation mechanisms:general - cosmology:theory 
\end{keywords}

% ************************************************************
\section{Introduction}
Lyman Alpha Emitters (LAEs) are galaxies identified by means of their strong Ly$\alpha$ emission. Unambiguous characteristics of this line including the strength, width, and blueward cut-off have enabled recent detections of LAEs out to $z \sim 7.7$ (Hibon et al. 2009), adding to the already existing data at $z\sim 7$ (Iye et al. 2006), $z\sim 6.5$ (Kashikawa et al. 2006), and lower redshifts. Since LAEs are among the most distant objects known, in addition to being superb probes of reionization, they serve as laboratories to study early galaxy evolution and clarify fundamental but poorly understood aspects of high-$z$ galaxies such as their dust content and extinction law. 

Although a number of studies have been conducted, the extent and sources of the dust enrichment of LAEs remains a much debated issue. Using cosmological SPH simulations, Dayal et al. (2009a) and Nagamine et al. (2008) have shown that at $z \sim 5.7$, the color excess of LAEs, $E(B-V) \sim 0.15$. These theoretical estimates are not inconsistent with recent experimental determinations: by fitting the SEDs of 3 LAEs at $z = 5.7$ by Lai et al. (2007) have inferred $E(B-V) < 0.225-0.425$; in a sample of 12 LAEs at $z = 4.5$, Finkelstein et al. (2009a) have found $A_{1200} = 0.5-4.5$; finally, Pirzkal et al. (2007) have found $A_V = 0.05-0.6$ for 3 galaxies at $z = 4-5.76$. The latter two values, when translated into the color excess (using a supernova dust extinction curve, e.g. Bianchi \& Schneider 2007) are found to be $E(B-V) = 0.035-0.316$ and $0.025-0.3$. These values therefore result in a coherent picture. However, Gronwall et al. (2007) and Ouchi et al. (2008) find much lower values of the color excess, $E(B-V) \leq 0.05$ for $z = 3.1$ LAEs.

The main uncertainty in using the SEDs to infer the dust content lies in the fact that the observed spectra depend on: (a) stellar population properties such as the age, metallicity, initial mass function (IMF) and (b) the amount and distribution of dust in the interstellar medium (ISM). These two effects are not easily disentangled from each other and this might be the source of the disparity in the dust extinction inferred by different works. These difficulties can be overcome if one can exploit the fact that dust absorbs ultraviolet (UV) radiation and re-emmits it at far-infrared (FIR) wavelengths; thus, if LAEs could be observed in the sum-millimeter (submm; FIR in the LAE's rest frame), a new window on the study of LAE properties would open, allowing to precisely determine their dust content and a far better determination of their stellar populations and star formation history.  

Clumped dust has also been invoked by many authors (Dawson et al. 2004; Dayal et al. 2008, 2009a, 2009b; Finkelstein et al. 2008, 2009a, 2009b; Kobayashi et al. 2007, 2009) to enhance the observed Ly$\alpha$ equivalent width (EW); the escape fraction of Ly$\alpha$ photons increases as compared to that of the continuum photons because of the so called `Neufeld effect' (Neufeld 1991), if dust in the ISM is clumped. If however, LAEs are found to contain negligible amounts of dust from FIR observations, scenarios which are considered exotic at present, including a top heavy IMF, PopIII stars and LAEs being extremely young objects (ages $\leq 10$ Myr) would have to be considered quite seriously. 

The main aim of this paper is to assess the detectability of LAEs in the submm band. We plan to accomplish this task by building upon a previously developed detailed model based on cosmological numerical simulations. Such model (Dayal et al 2009b) successfully predicts the Ly$\alpha$ and UV Luminosity Function (LF) of the LAE galaxy population; in addition, it provides a wealth of information about the physical properties of these objects.

% **********************************************************
\section{Method}
\subsection{Basic simulations}\label{basic sim}
We start the calculation from a previously developed model of LAEs, the details of which can be found in Dayal et al. (2009b). In brief,  the cosmological SPH simulations\footnote{The adopted cosmological model for the simulation corresponds to the $\Lambda$CDM Universe with $\Omega_{\rm m }=0.26, \Omega_{\Lambda}=0.74,\ \Omega_{\rm b}=0.0413$, $n_s=0.95$, $H_0 = 73$ km s$^{-1}$ Mpc$^{-1}$ and $\sigma_8=0.8$, consistent with the 5-year analysis of the WMAP data (Komatsu et al. 2009).} of $(75 h^{-1}{\rm Mpc})^3$ (comoving) have been carried out using a TreePM-SPH code 
GADGET-2 (Springel 2005) with the implementation of chemodynamics as described in Tornatore et al. (2007). Galaxies are recognized as gravitationally bound groups of star particles by running a standard friends-of-friends (FOF) algorithm, decomposing each FOF group into a set of disjoint substructures and identifying these. After performing a gravitational unbinding procedure, only sub-halos with at least 20 bound particles are considered to be genuine structures, Saro et al. 2006). For each ``bona-fide'' galaxy, we compute the mass-weighted age, the total halo/stellar/gas mass, the star formation rate (SFR), the mass weighted gas/stellar metallicity, the mass-weighted gas temperature and the half mass radius of the dark matter halo. 

We have coupled  the above simulations with a Ly$\alpha$/ continuum production/ transmission model as in Dayal et al. (2009a). For each galaxy, the total Ly$\alpha$ and continuum luminosity are calculated as a sum of the contributions from (a) stellar processes and (b) cooling of collisionally excited \HI in the ISM (Dayal et al. 2009b). In the post-processing, assuming Type II supernovae (SNII) to be the primary dust factories for $z \geq 5.7$, we include a model to calculate the total amount of dust in each galaxy depending on its intrinsic properties such as the SFR, age, metallicity and gas mass. This dust mass is then used to calculate the associated optical depth, which is translated into the escape fraction for continuum photons, $f_c$, assuming a slab-like dust distribution, to obtain the UV luminosity function (LF). The escape fraction of Ly$\alpha$ photons is then assumed to scale with $f_c$ to obtain 
the observed Ly$\alpha$ LF. 

Galaxies are identified as LAEs based on the current operational definition, i.e. observed Ly$\alpha$ luminosity, $L_\alpha \geq 10^{42.2}\, {\rm erg\, s^{-1}}$ and observed equivalent width, $EW \geq 20$ \AA. In total, (39000,18000) galaxies have been identified in the two relevant simulation outputs at $z = (5.7, 6.6)$ out of which (1045, 502) have been selected as LAEs according to the above criterion. 

% ***************************************************************************
\subsection{UV escape fraction and FIR luminosity}

We estimate the total FIR luminosity for all the galaxies identified as LAEs at the redshifts of interest, i.e. $z \sim (5.7,6.6)$. For this calculation, we assume that the dust is predominantly heated by UV radiation from stars with wavelengths 912--4000 \AA\,, Buat \& Xu (1996); in this paper, the term UV is used for non-ionizing continuum in this wavelength range. We start by calculating the optical depth to dust, $\tau_{UV}$, seen by UV photons as 
$\tau_{UV} = {3 \Sigma_{d}}{[4as]^{-1}}$, where $\Sigma_{d}=M_{ dust}/(\pi r_d^{2})$ is the dust surface mass density, $a$ and $s$ are the grain radius and material density. The dust surface mass density is estimated by assuming that the dust mass, $M_{dust}$ is spread on a scale, $r_d= (0.6,1.0)r_e$ at $z \sim (5.7,6.6)$, where $r_e$ is the stellar distribution scale inferred using the results of Bolton et al. (2008). Complete details of this 
calculation can be found in Dayal et al. (2009b).

Assuming the dust and stars to be homogeneously distributed, we use this optical depth to calculate the UV luminosity, $L_{UV}$, escaping from the galaxy as $L_{UV} = L_{UV}^{0}f_c,$ where $L_{UV}^0$ is the intrinsic UV luminosity, calculated using the population synthesis code {\tt STARBURST99} (Leitherer 1999) using the appropriate IMF, star formation rate ($\dot M_*$), metallicity ($Z$) and age ($t_*$) for each LAE, as obtained from the simulation. Further, $f_c$, the escape fraction of the UV continuum is calculated assuming a slab-like dust distribution, such that
\begin{eqnarray}
f_\mathrm{c}=
\frac{1-e^{-\tau_{UV}}}{\tau_{UV}}.
\label{fc}
\end{eqnarray}
Hence, $f_c$ depends on the intrinsic properties of the galaxies, i.e, the dust mass and dust distribution scale. The average value of $\langle f_c \rangle=(0.23,0.38)$ at $z \sim (5.7,6.6)$ respectively, Dayal et al.\ 2009b). 

Because of the large cross section of dust against UV light and the intense UV field in a star forming galaxy, we can assume that the emitted FIR luminosity, $L_{FIR}$, is equal to the UV luminosity which is absorbed and heats up the dust, such that 
\begin{equation}
L_{FIR} = L_{{UV}}^{0} - L_{{UV}} = (1-f_c)L_{{UV}}^{0}.
\end{equation}

In order to examine the detectability of dust emission, the observed flux is predicted to be
\begin{eqnarray}
f_\nu =\frac{(1+z)L_{\nu (1+z)}}{4\pi d_{L}^2},
\end{eqnarray}
where $d_{L}$ is the luminosity distance (Carroll, Press \& Turner 1992), and $L_\nu$ is the monochromatic luminosity at the chosen frequency of observation.
% ($3.5\times 10^{13}$ Hz) corresponding to a wavelength of $850 \mu$m for this work

If dust grains emit thermally with a single dust temperature, $T_{dust}$, $L_\nu$ can be written as
\begin{equation}
L_\nu  =  4\pi M_{dust}\kappa_\nu B_\nu (T_{dust}),
\label{lnu}
\end{equation}
where $\kappa_\nu$ is the mass absorption coefficient, $B_\nu (T_{dust})$ is the Planck function with frequency $\nu$ and temperature $T_{dust}$.

If we assume a power law form for $\kappa_\nu$ such that $\kappa_\nu = \kappa_{\nu_0} (\nu /\nu_0)^\beta$, the total FIR luminosity can be expressed as 
\begin{eqnarray}
L_{FIR} & \hspace{-2mm}= & \hspace{-2mm}\int_0^\infty L_\nu\,d\nu\nonumber\\
& & \hspace{-13mm}=4\pi M_{dust}\kappa_{\nu_0}
\nu_0^{-\beta}\left(\frac{kT_{dust}}{h}\right)^{4+\beta} \left(\frac{2h}{c^2}\right)
\int_0^\infty\frac{x^{3+\beta}}{e^x-1}dx,
\label{Lnu_integ}
\end{eqnarray}
where, $k$ is the Boltzmann constant, $h$ is the Planck constant, $c$ is the light speed and $x = h\nu(k T_{dust})^{-1}$. We use $\nu_0 = 3.00\times 10^{12}$ Hz (corresponding to a wavelength of 100 $\mu$m), which is an arbitrary frequency in FIR used for normalization.

Since we assume all the dust grains to be spherical with a single size $a = 0.05 \mu$m (appropriate for a SN-produced dust, see Todini \& Ferrara 2001; Nozawa 2003, 2007) and material density $s$, the mass absorption coefficient $\kappa_{\nu_0}$ can be written as $\kappa_{\nu_0} =3Q_\nu[4as]^{-1},$ where $Q_\nu$ is the optical absorption cross section normalized to the geometrical cross section ($\pi a^2$). For graphite/carbonaceous grains, $Q_\nu\,  a^{-1}=1.57\times 10^2$ cm$^{-1}$ at a wavelength of 100 $\mu$m, $s=2.25$ g cm$^{-3}$ and $\beta =2$ (Draine \& Lee 1984). This results in $\kappa_{\nu}=52.2\, (\nu /\nu_0)^\beta$ cm$^2$ g$^{-1}$.
\footnote{The corresponding values for silicates are $Q_\nu \, a^{-1}=1.45\times 10^2$ cm$^{-1}$, $s=3.3$ g cm$^{-3}$ and $\beta =2$, which results in $\kappa_{\nu}=32.9\, (\nu /\nu_0)^\beta$ cm$^2$ g$^{-1}$.} Even though in this work, we assume the dust grains to be graphites, the flux is insensitive to the specific grain material for a given value of $L_{FIR}$ since a small $\kappa_\nu$ is compensated by a high $T_{dust}$ and vice versa. Then, equation (\ref{Lnu_integ}) can be simplified to the following expression
\begin{eqnarray}
T_\mathrm{dust}= 6.73 \left(
\frac{L_\mathrm{FIR}/L_\odot}{M_\mathrm{dust}/M_\odot}
\right)^{1/6}~\mathrm{K}.
\end{eqnarray} 
This temperature is put in eq. (\ref{lnu}) to obtain $L_\nu$ and hence the dust emission flux detectable in the submm bands in the observer's frame. 

% ************************************************************************
\section{Results}

\subsection{Dust abundance}
\label{dust abundance}

The relation between dust-to-gas ratio ($D$) and gas metallicity is useful to discuss the dust enrichment scenario in the low-metallicity phase as shown by Lisenfeld \& Ferrara (1998) for nearby blue compact dwarf galaxies. We calculate this ratio as $D = M_{dust}/M_g$, where the dust mass, $M_{dust}$ is calculated as shown in Dayal et al. (2009b). The gas mass, $M_g$ and the mass-weighted gas metallicity, $Z_g$, are both obtained from the simulation as mentioned before. We normalize the dust-to-gas ratio to the one measured in the  solar neighborhood, $D_\odot= 1/150 = 6\times 10^{-3}$ (Hirashita \& Ferrara 2002).  
\begin{figure} 
%  \vspace*{10pt} 
  \center{\includegraphics[scale=0.5]{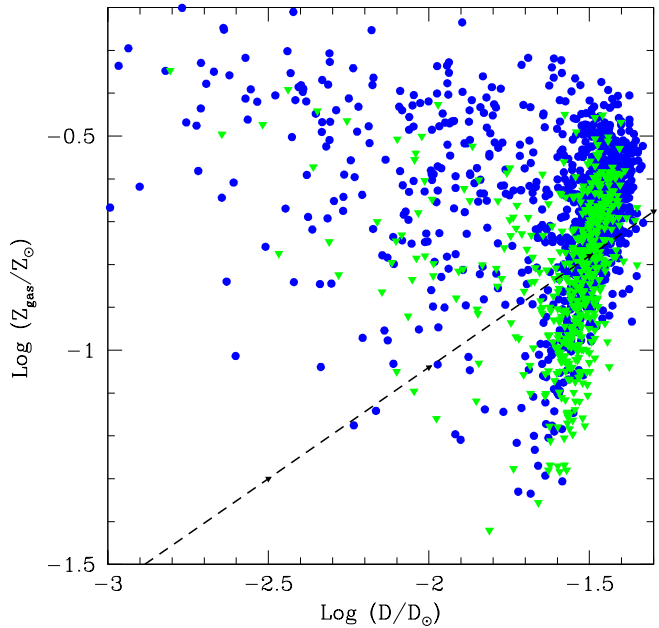}} 
  \caption{Relation between dust-to-gas ratio and gas metallicity for LAEs at $z=5.7$ (circles), 6.6 (triangles). The dashed line corresponds to $Z_g \propto {D}^{0.52}$ (Lisenfeld \& Ferrara 1998). }
\label{dust_gas} 
\end{figure} 
The relation between $D$ and $Z_g$ has been investigated by several authors. In particular, Lisenfeld \& Ferrara (1998) have shown that for nearby dwarf galaxies the following relation holds: $Z_g \propto {D}^p$, with $p=0.52\pm 0.25$. In principle, there is no reason to expect that such low-redshift determination applies equally well to LAEs, due to the very different physical conditions and shorter evolutionary timescales allowed by the Hubble time at $z\approx 6$. Using the above relation for a gas metallicity value equal to $0.2 Z_\odot$, appropriate for our calculation of LAEs, yields $D = 0.003-0.12 D_\odot$; from our result we get a mean value in reasonable agreement with this expectation, $D = 0.001-0.045 D_\odot$ for $Z_g = 0.2 Z_\odot$. 

However, although we confirm a correlation between dust abundance and metallicity, the $D-Z_g$ relation we find, shown in Fig. 1, shows considerable deviations from the above power-law. First, the slope of the relation is considerably steeper (i.e. $p >1$); second, in spite of the relatively high metallicity of LAES (about $0.1-0.5 Z_\odot$), the dust abundance is relatively depressed with essentially all objects having $D < 0.05 D_\odot$. Taken together, these two points imply that LAEs are relatively poorly efficient dust producers with respect to local galaxies. The physical explanation of this fact is straightforward. Because of their young ages ($< 500$~Myr), LAEs can only have their dust produced by SNII rather than by evolved stars, which dominate the dust production in local galaxies. However, since dust is both produced and destroyed by SN in LAEs, $D$ saturates at a relatively low value; in local galaxies, instead, SN shocks cannot counteract the rate of dust production by evolved stars and $D$ grows to 
larger values.   

\begin{figure*} 
%  \vspace*{10pt} 
  \center{\includegraphics[scale=1.0]{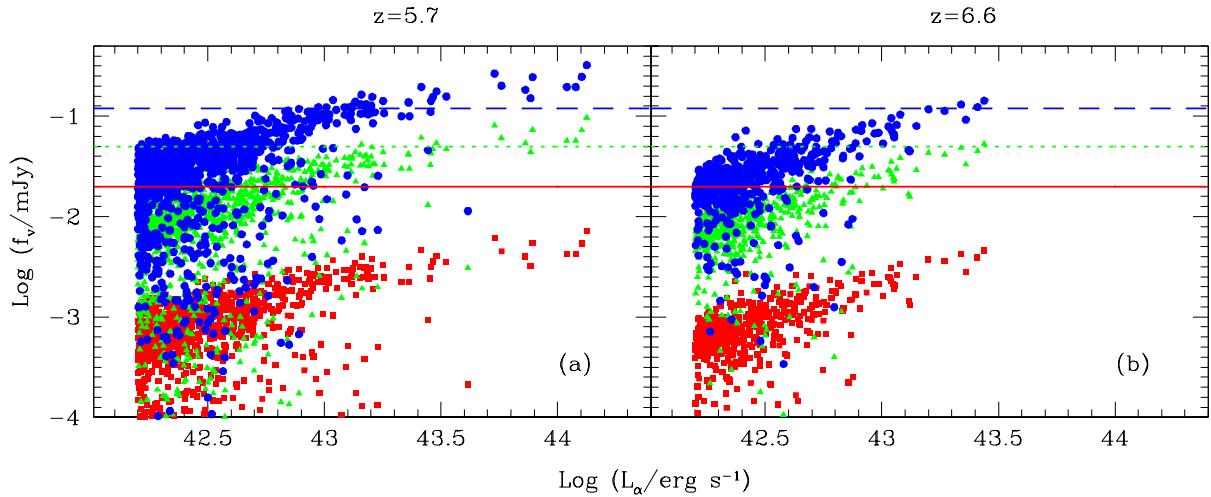}} 
 \caption{Correlations between Ly$\alpha$ luminosity and submm flux for $z=5.7$ (left) and $6.6$ (right). The horizontal solid, dotted and dashed lines show the $5\sigma$ detection limit of ALMA with 1 hour integration for the bands corresponding to $100$ Ghz (3mm), $220$ Ghz (1.4 mm) and $353$ GHz (850$\mu$m) respectively. Symbols (squares, triangles, circles) in both panels show the results obtained from this work for the same bands (100, 220, 353 GHz).}
\label{alma} 
\end{figure*}

Finally, Fig. 1 shows the presence of a considerable number of outliers to the left of the power-law relation, i.e. the low $D$-high  $Z_g$ region, at both redshifts. Since both the dust and gas mass scale with the halo mass (Dayal et al. 2009b), this physically means that even some galaxies with a very small halo mass retain a large metal fraction in the ISM gas. This can possibly be explained by the fact that though galaxies with a small halo mass have very feeble star formation rates and hence produce less amount of metals as compared to larger galaxies, the mechanical feedback that would eject their ISM metal content is comparatively more depressed with respect to larger galaxies.
 
% ********************************************************
\subsection{Prediction for sub-millimeter fluxes}

ALMA could be a strong tool to directly detect the dust emission from a large number of LAEs. Then, the question of how to identify the sources of submm emission arises. We find that the observed Ly$\alpha$ luminosity at both $z \sim (5.7,6.6)$ (as calculated in Dayal et al. 2009b) and the fluxes at $3$mm, $1.4$ mm and $850 \mu$m of LAEs are correlated, as shown in Fig. \ref{alma}; in general, the galaxies with the largest observed Ly$\alpha$ luminosities also show the largest value of the FIR fluxes. The reason for this can be explained as follows: the observed Ly$\alpha$ luminosity, $L_\alpha$ is related to the intrinsic value, $L_\alpha^{int}$ as
\begin{equation}
L_\alpha = L_\alpha^{int} f_\alpha T_\alpha,
\end{equation}
where $f_\alpha$ is the fraction of Ly$\alpha$ luminosity which emerges from the galaxy, undamped by dust and $T_\alpha$ is the fraction of Ly$\alpha$ luminosity transmitted through the IGM. Now, the largest galaxies have the 
largest star formation rates (Dayal et al. 2009a) and hence the largest values of $L_\alpha^{int}$ since this scales with the star formation rate. As shown in Dayal et al. (2009b), their $T_\alpha$ is also the largest. Further, the largest galaxies are also the most dust-enriched by SNII since the SNII rate scales with the star formation rate. However, due to large radii of dust distribution (which scales with the SFR), the largest galaxies do not have the smallest $f_\alpha$ as expected. This implies that the largest galaxies show both the largest $L_\alpha$ and $f_\nu$, hence the trend seen. Fig. \ref{alma} shows that  the $850 \mu$m band is the optimum one to look for submm emission; (32,\,5) LAEs at $z=(5.7,\,6.6)$ in our simulation volume have an 850$\mu$m flux larger than the $5\sigma$ 1 hour integration limit of ALMA which is $0.12$ mJy (Morita \& Holdaway 2005). This corresponds to about $(3\%,1\%)$  of 
the LAEs in our simulated samples at these redshifts.

\begin{figure} 
%  \vspace*{10pt} 
  \center{\includegraphics[scale=0.5]{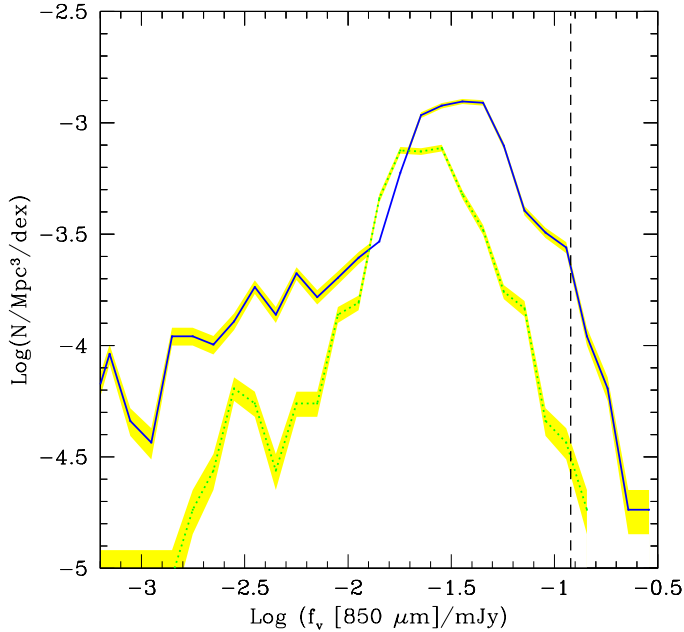}} 
  \caption{The distribution function of the observed $850\mu$m flux of LAEs at $z=5.7$ (solid line) and $6.6$ (dotted line). The vertical dashed line shows the $5\sigma$ 1 hour integration detection limit of ALMA ($=0.12$ mJy) and the shaded regions show the poissonian errors.}
\label{lf} 
\end{figure}

In Fig. \ref{lf}, we show the FIR luminosity and submm flux luminosity functions for our LAE sample at $850 \mu$m since this seems to be the most optimum band to look for FIR emission, as mentioned above. We see that selecting LAEs with $L_\alpha \geq 10^{42.2} {\rm erg\, s^{-1}}$, can efficiently detect galaxies whose submm fluxes are bright enough to be detected by ALMA with an integration time of only 1 hour. A longer integration time drastically increases the number of LAEs detected by ALMA because of the steep rise of distribution function toward faint submm fluxes as shown in Fig. \ref{lf}.

Thus, we show that we can simply point ALMA towards bright LAEs that have already been identified at high-z to detect LAEs in the sub-mm band at $5\sigma$ with an integration time of only 1 hour.

%\texttt{I would say that a survey of a cosmological volume like (100 Mpc)$^3$ is unrealistic for ALMA. 100 Mpc corresponds to 4.3 deg at z = 5. So we have to survey like 16 deg$^2$ = 2.1 x 10$^8$ arcsec$^2$. Since the field of view of ALMA at 850 um is as small as $\sim 60$ arcsec$^2$, 3.5 x 10$^6$ pointings are necessary. If we take 1 hour per pointing, we require 3.5 x 10$^6$ hours, which is 400 years. So I would recommend to discuss only the pointing observations of each LAEs.}

% *****************************************************************
\section{Summary and conclusions}
This work is based on the results obtained in Dayal et al. (2009b), where, we coupled a cosmological SPH simulation with a Ly$\alpha$/continuum production/transmission model and a dust model. This enabled us to calculate the observed Ly$\alpha$/continuum luminosities and the dust mass evolution for each galaxy. We thus obtained the escape fraction of continuum photons for each galaxy based on its intrinsic properties including the star formation rate, age, metallicity, IMF and gas mass. The only free parameter, the escape fraction of Ly$\alpha$ photons was obtained by comparing the model results to the observations at $z \sim (5.7,6.6)$. 

Though we find that the dust to gas ratio values as a function of gas metallicity obtained from our model are broadly consistent with those predicted by Lisenfeld \& Ferrara (1998) at $Z =0.2 Z_\odot$, the slopes predicted by the two works are different. The reason for this disagreement arises because we assume SNII to be the only sources of dust. While this leads to a progressive metal enrichment of the ISM and hence a high value of the gas metallicity, since SNII also destroy the dust that they produce, the dust to gas ratio remains very small ($< 0.05$ of that in the solar neighborhood). On the other hand, in dwarf galaxies, both SNII and evolved stars produce dust; the supernovae shocks, however, are not capable of destroying the huge amounts of dust produced by evolved stars and this leads to much larger values of the dust to gas ratio.

For each galaxy identified as a LAE as these redshifts, we assume that the UV luminosity between 912-4000 \AA\, that is absorbed inside the galaxy heats up the dust, leading to re-radiation in the submm band in the observer's frame. This allows us to calculate the emitted flux at 3000, 1363 and 850 $\mu$m. 

We then present a very efficient strategy to look for dust emission from LAEs. We find that the observed Ly$\alpha$ luminosity and the FIR fluxes are correlated for LAEs in all the three ALMA bands (3000, 1363 and 850 $\mu$m); the LAEs with the largest observed Ly$\alpha$ luminosity also show the largest fluxes. This means that pointing surveys of the brightest LAEs already identified would yield large fluxes in ALMA. We find that the $850 \mu$m band seems to be the optimum one for detecting the FIR emission from high-z LAEs. Although a few tens of objects would be detectable with a 1 hour integration, the number density of objects detectable would rise steeply with the integration time.

Finkelstein et al. (2009b) have used the measurements of the UV spectral slopes to derive far-infrared flux predictions for a sample of 23 LAEs (Finkelstein et al. 2009a; Pirzkal et al. 
2007) at z$\geq 4$. It is interesting to compare to what extent the results obtained using their approach match with those from our model. We first compare the color excess obtained from these works where we find that the average value, $E(B-V)\sim0.15$ obtained from our work (Dayal et al. 2009a; 2009b) is very consistent with $E(B-V)=0.035-0.316$ and $0.025-0.3$, found by Finkelstein et al. (2009a) and Pirzkal et al. (2007) respectively. However, while Finkelstein et al. (2009b) find that $39\pm22\%$ of their LAEs would be visible with ALMA with an integration time of 4 hours, we find only $\sim (3\%,1\%)$ of the LAEs in our simulation would be visible with a $5-\sigma$, 1 hour integration limit at $z\sim(5.7,6.6)$. Although these results may seem discrepant at the first glance, they can be explained quite easily considering the redshifts of the LAEs modelled by Finkelstein et al. (2009b). In their sample, 20 of the LAEs are at $4\leq z \leq 5$, of which, 9 (i.e. 45\%)  would be visible with ALMA with a 4 hour integration time. However, \textit {none} of the 3 LAEs with $z>5.2$ would be detectable in the submm. This is highly consistent with our results, where only 3\% of the LAEs would be visible with ALMA at $z\sim 5.7$ for a $5-\sigma$ 1 hour integration time; hence, in a sample of 3 LAEs, none would be visible. This is also a confirmation of our dust model, where SNII are the primary dust factories at $z \geq 5.7$, with evolved stars dominating at lower redshifts. Since SNII both produce and destroy dust, the dust enrichment levels at high redshifts are very small as compared to that at lower redshifts ($z \leq 5.7$). Hence, the detectability of LAEs in the submm decreases with increasing redshift. 

The main uncertainty in this work lies in the fact that we assume SNII to be the primary sources of dust production/destruction at the redshifts considered. While this is justified for ages $\leq 1$ Gyr, we can not completely rule out the contribution from evolved stars as shown by Valiante et al. (2009). Further, the dust yield per SNII and the dust destruction efficiency due to supernova shocks are only known well enough to within a factor of a few. Also, since the scale length of dust distribution is not well known at these redshifts, we have to resort to the data to guide us. We plan to deal with these uncertainties in future works.

% ***************************************************************************
\section*{Acknowledgments} 
We thank the anonymous referee for valuable comments which have contributed positively to the presentation and readability of the paper. PD thanks A. Mazumdar for interesting discussions.

% **************************************************************************

%\bibitem[\protect\citeauthoryear{}{}]{b} 

\newpage 
\label{lastpage} 
\end{document}